\nonstopmode
\documentclass[preprint,showpacs,preprintnumbers,amsmath,amssymb,aps]{revtex4}
\usepackage{textcomp}
\input boxedeps 
\input epsf
\begin{document}
\title{Free Surface Deformation and Cusp Formation During the Drainage of a Very Viscous Fluid}

\author{Sahraoui Cha\"{\i}eb\footnote{email: sch@uiuc.edu}}

\affiliation{ Department of 
Theoretical and Applied Mechanics, University of Illinois at Urbana-Champaign, Urbana 
IL 61801 USA.}

\vskip .2cm
\begin{abstract}
We report an experimental study of the deformation of  the free surface of a very viscous fluid when drained out of a container through a circular orifice. At some critical height of the fluid, the free surface
deforms and forms a dimple. At later times, this dimple becomes a cusp. We found that the height of the dimple measured from the bottom of the tank as well as the curvature of the dimple before the cusp formation, follows power laws behaviors. We found that this behavior was due to an interplay between viscous pressure and surface tension. 
\end{abstract}

\maketitle
\vskip .2cm

When a normal fluid, such as water, is drained out of a container such as a bath tub the free surface deforms and, if inertia dominates,
a whirlpool forms. This whirlpool is in the form of a vortex or a dimple
 which apex will be at the level of the bottom surface
of the tank just above the draining hole \cite{springer}. 
 We have noticed that if the liquid we drain is very viscous, or in other words, when viscous effects dominate the flow, 
the dynamics and the form of the deformed free surface are no 
 longer the same. The dimple appears far away from the drainage hole in the vertical direction, and 
it evolves into a sharp dimple at early stages of the draining.
In this paper, we report such effect where the free surface of a very viscous fluid 
deforms into a  dimple  when drained from a cylindrical container. At later times this dimple evolves into a cusp. 
Such free surface deformations are also found in selective withdrawal\cite{hocking,blake, lister,itai1,brenner,eggers1,dupont}, where the flow is steady whereas our flow is gravity driven and non-steady. We, however,  fitted the profile of the dimple to a power law of the $z \propto x^\alpha$  and found that the power $\alpha $ is a linear
function of time. This might be an effect of the non-steadiness of the interface motion.
 
\begin{figure}[tbh]
\centerline{\epsfxsize=\columnwidth \epsfysize=.75\columnwidth 
\epsfbox{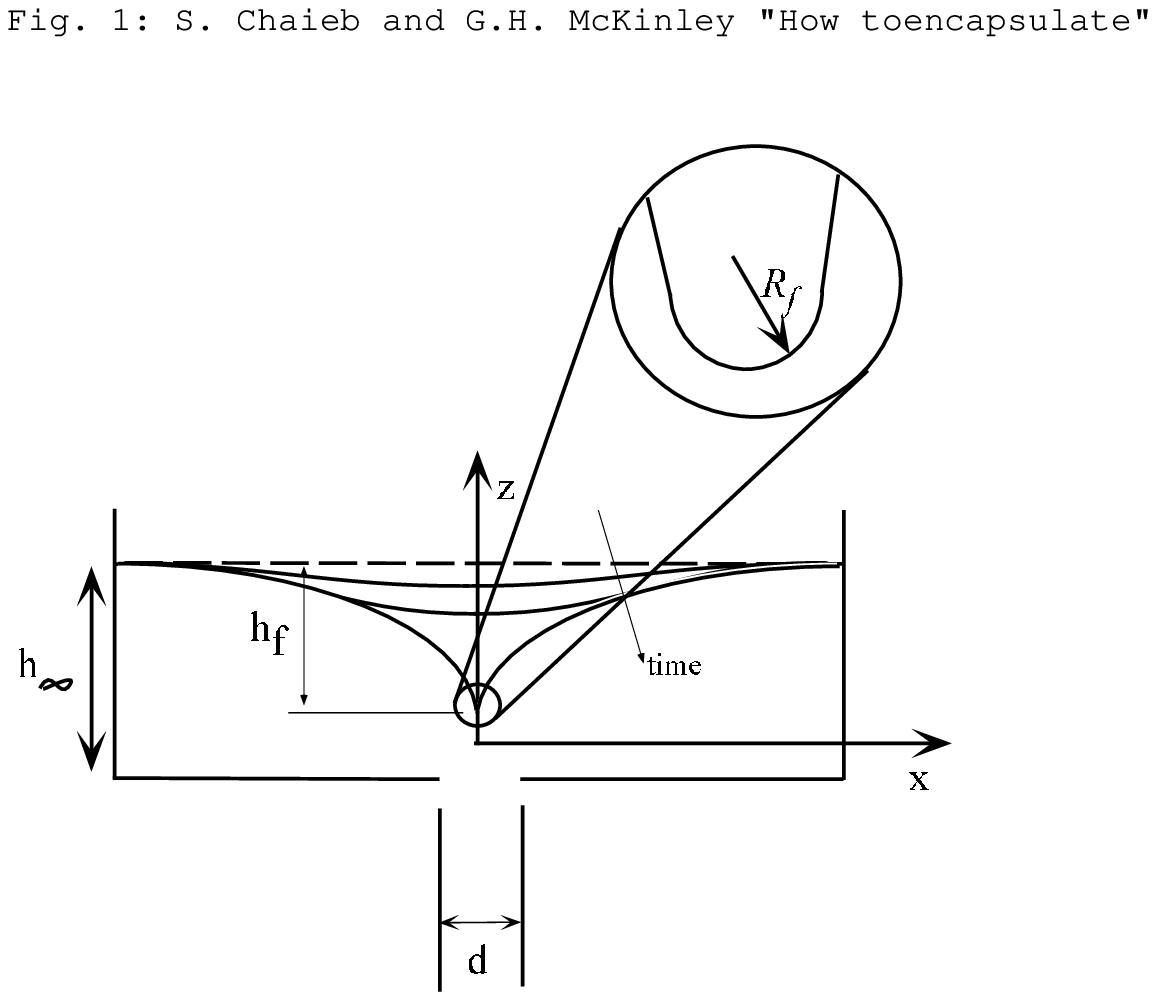}}
\caption{The sketch of the set up and the definition of $\cal{R}$ and $h_f$ which 
is the radius of curvature of the cusped surface and the height of the 
hollow dimple that evolves to a cusp respectively.}
\label{setup}
\end{figure}

A scheme of the set up and the axis notation is displayed in 
figure~\ref{setup}. 
We used viscous liquids (from the PDMS family: called also silicon oils) which viscosity ranges from 
10 to 1000 Pa.s. (viscosity of water is 0.001 Pa. s.). The fluids are drained out of the container under
the effect of gravity through an axisymmetrically placed circular 
orifice of diameter ranging from 1 to 4 mm at the bottom of a 107 mm diameter cylindrical container.

\begin{figure}[htb!]

\centerline{\epsfxsize=0.33\columnwidth\epsfysize=.33\columnwidth
\epsfbox{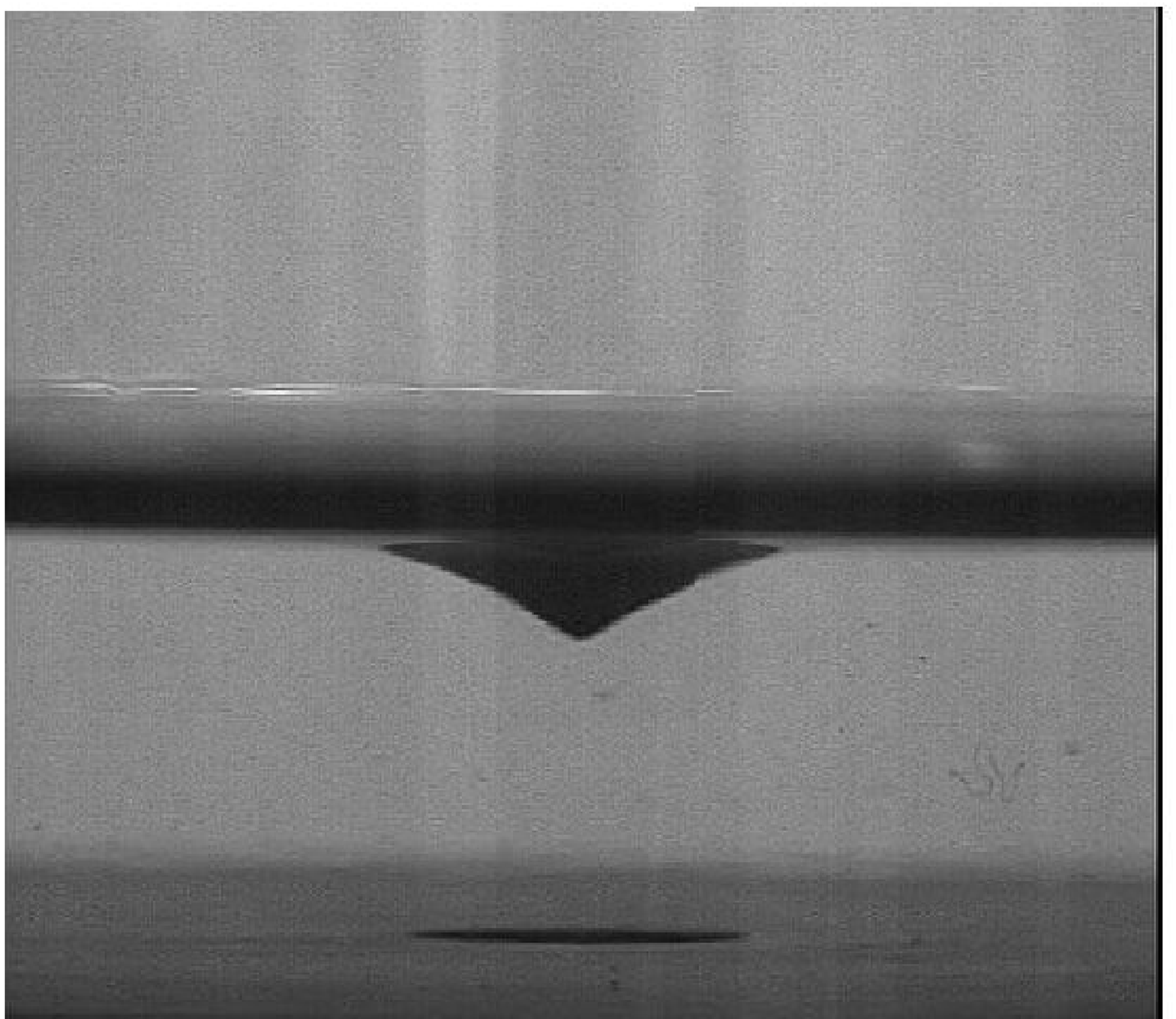}
\epsfxsize=0.33\columnwidth\epsfysize=.33\columnwidth\epsfbox{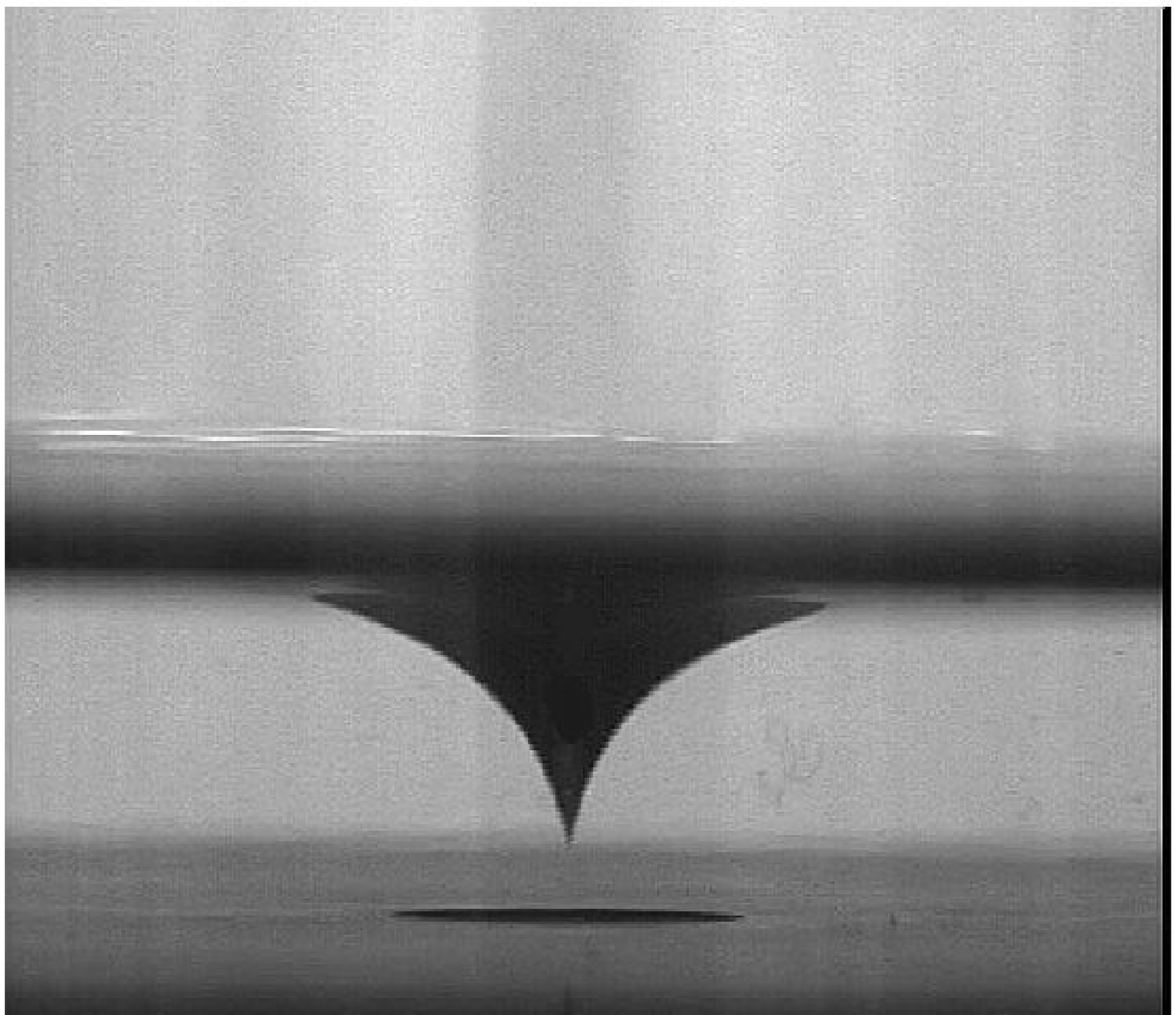}
\epsfxsize=0.33\columnwidth\epsfysize=.33\columnwidth\epsfbox{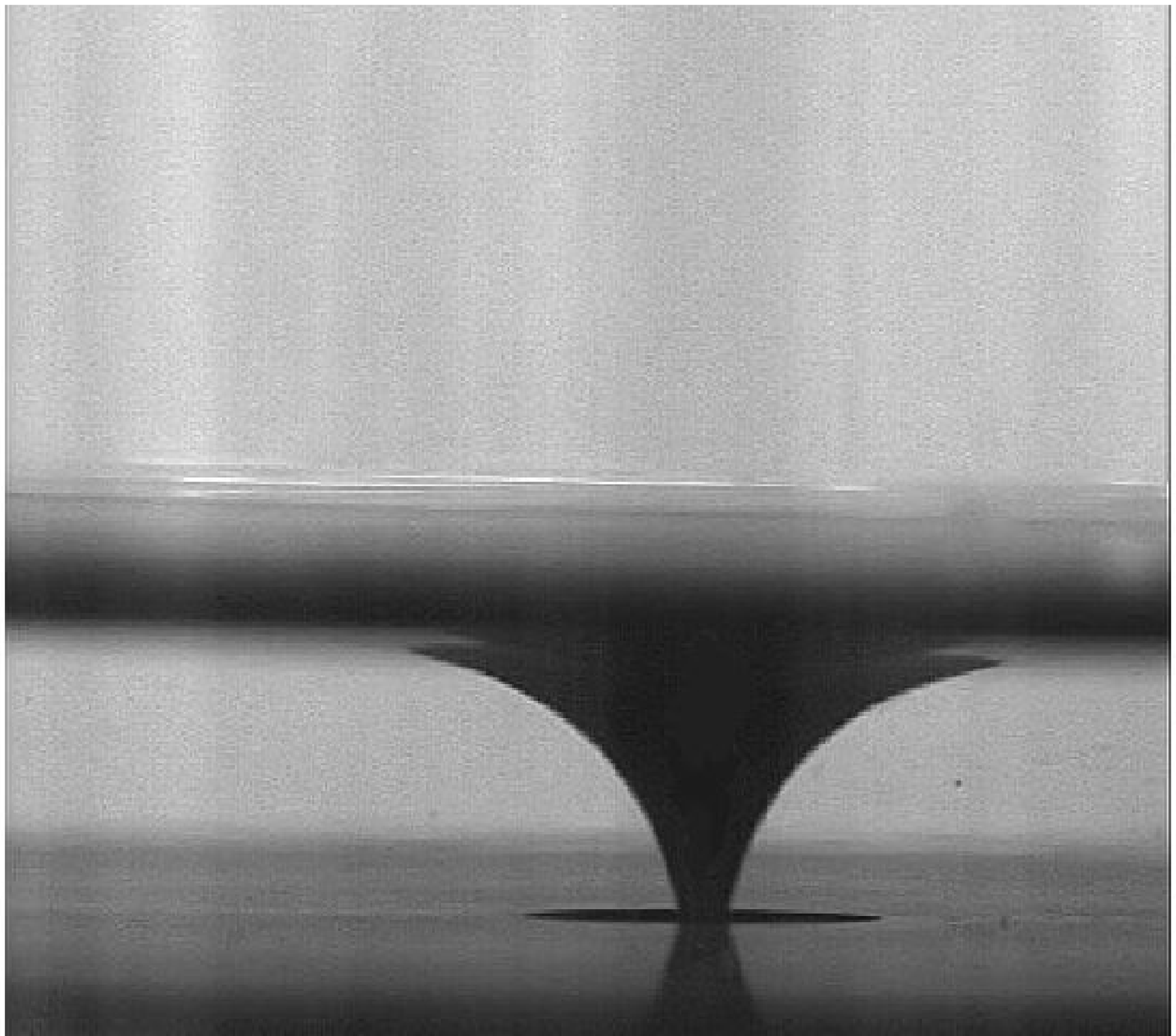}}
\caption{ Sketch of the time series of the drainage process. In 
{\bf{\textsf{a}}}, we display early time of the drainage process. Notice that 
the surface is smooth and the dimple is not singular. In {\bf 
{\textsf{b}}},  
Intermediate time in the draining process. Notice the singular aspect of 
the dimple. This picture was taken just before the break-up into a thread 
of droplets. 
In picture {\bf{\textsf{c}}}, the tip of the dimple is now entering the hole and is 
following the thread which is a hollow tube where air is being 
encapsulated. Another liquid can be encapsulated too (later in the text.). 
The viscosity of the liquid used in these pictures is $10^3$ Pa. s.}
\label{cusp}
\end{figure}

As the liquid drains, the liquid-air interface dimples above the drain hole.
In figure~\ref{cusp}, we show a sequence of how the smooth dimple evolves in
time to become cusp-like. Notice the transition from a smooth dimple in 
figure~\ref{cusp}a to a sharp interface in figure~\ref{cusp}b. In 
figure~\ref{cusp}c the free surface enters the hole and the bottom of the 
cusp is far below the bottom of the container. 
As the level of the liquid falls during the liquid drainage, the dimple becomes sharper. 
The dimple in the 
liquid-air interface breaks up at some critical height $h_f$ measured from the bottom
surface of the container and begins to entrain a thread of air. The moment where the thread of 
air bubbles appear will be called $t_f$ and will be associated to $h_f$. 
In other words, at $h_f$ the interface undergoes a topological transition 
in a manner analogous to the pinch-off of a viscous thread. At the 
transition, the curvature $\kappa$ of the interface at the dimple becomes 
infinite, i.e. a cusp forms. 

\begin{figure}[htb!]
\centerline{\epsfysize=\columnwidth \epsfbox{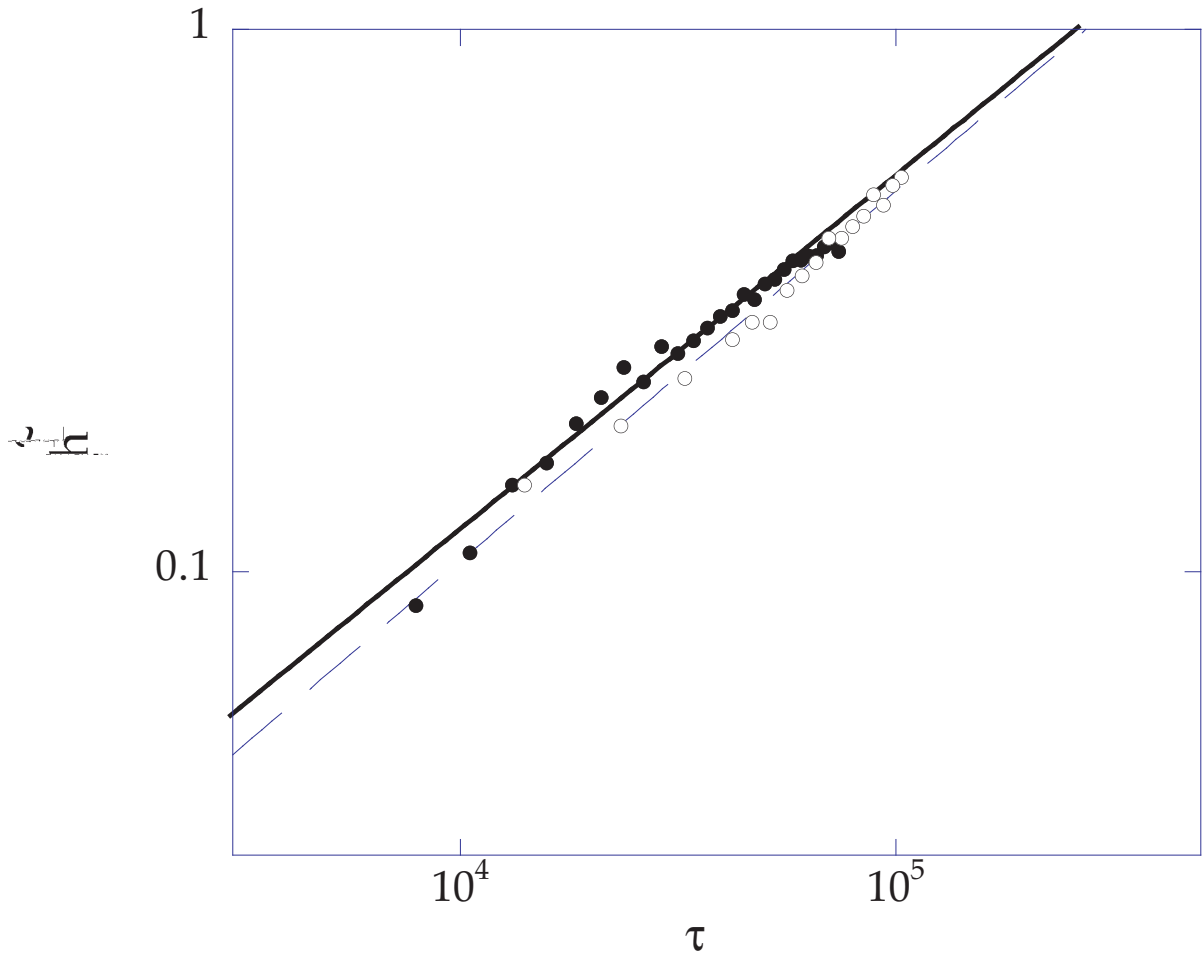}}
\caption{The normalized height of the dimpled region just before ``cusping''
versus the normalized time scale. The closed circles correspond to viscosity of $10^5$ cSt and 
the open circles correspond to a viscosity of $10^4$ cSt. The data are  
best fit by a 2/3 power law.}.
\label{height}
\end{figure}

We measured the height of the dimple $h_f$ and its corresponding time $t_f$ with an error of 1 mm and 1 sec respectively. This is checked by looking at the video frame just before some air bubbles start nucleating at the dimple's apex. We use standard video with a frame frequency of
30 frames per second. 
We define the non dimensional number $\tilde{h}=\frac{h-h_f}{D}$ where 
$D$ is the hole diameter as shown in figure ~\ref{setup}. The flow 
rate is varied by varying the hole diameter.
We measured the dependence of $\tilde{h}$
versus the dimensionless time scale defined
in our case as ${\mathsf{\tau}}=\frac{\nu(t-t_f)}{ D^2}$.  In this expression
$\nu$ is the kinematic viscosity  $\mu/\rho$. The result is displayed in 
figure~\ref{height}. The Log-Log plot shows a power law dependence and the line is
 best fit by a 2/3 power law: $\tilde{h} \sim {\tau}^{2/3}$. 

We monitored, $ h_{\infty }$,which is the height of the liquid far away from the cusp region
as shown in Fig.~\ref{setup}, versus time and found the flow rate which gives 
the velocity at the draining hole.  The Reynolds number defined as
$\frac{\rho U R}{\mu}$ is equal to $3.3\times 10^{-3}$ for the liquid of viscosity 10 Pa. s.  and is
equal to $ 2.7 \times 10^{-6}$ for a liquid of viscosity 100 Pa.s. Where 
$\rho$ is the density of the liquid, $U$ is the flow velocity of the liquid at the hole and $R$ is the exit-hole radius

\begin{figure}[htb!]
\centerline{\epsfysize=\columnwidth \epsfbox{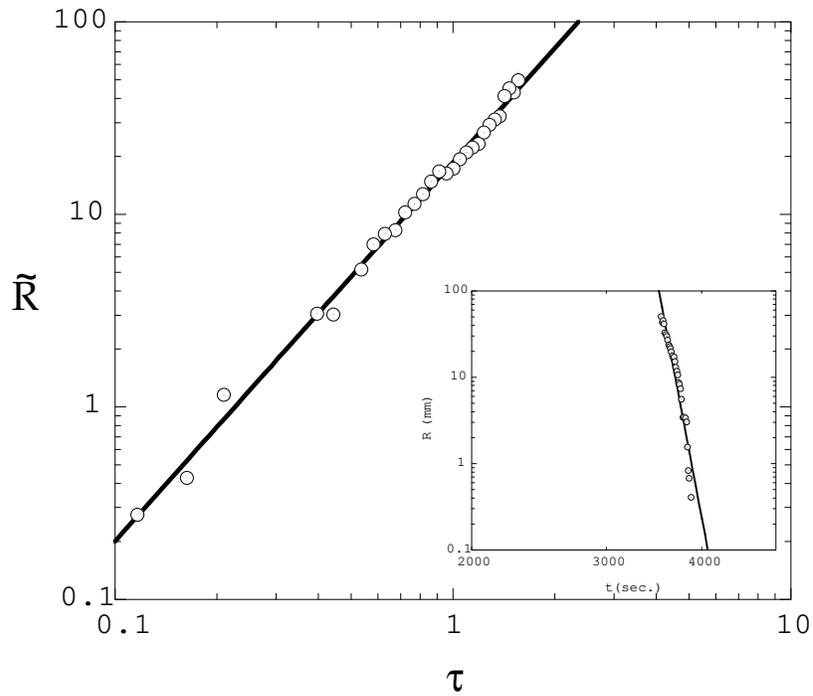}}
\caption{The normalized radius of curvature of the dimple, as defined in 
figure~\ref{setup}, versus the normalized time scale.
 The line is the best fit of the data to the power 2. Inset: The drop of the radius of 
 curvature versus time. }
\label{curvature}
\end{figure}

This cusp formation is analogous to the selective withdrawal and to drop pinch off.
Here the critical dimple height $h_f$ is analogous to the point
$z_0$ for thread pinch-off \cite{eggers1}. The distance between 
the current dimple height and $h_f$, $dh=h-h_f$, is analogous to the 
minimal thread radius for thread pinch-off \cite{itai1,brenner}.
As the interface's dimple sharpens to a cusp, the capillary stress 
$\gamma\kappa$ diverges.
To check this ideas we measured the radius of curvature $\mathcal{R}$ of the
 dimple, by fitting it to a polynomial, while it is still smooth before the appearance of the bubbles, 
defined in the magnified area of figure~\ref{setup}. The plot in figure~\ref{curvature} is the value of the curvature $\tilde{\mathcal{R}}= ({\mathcal{R- R_{\text{f}}}})/D$ as defined in 
fig~\ref{setup}   versus the quantity $\tau$. Similarly 
${\mathcal{R_{\text{f}}}}$ is the radius of curvature just before the nucleation of 
the bubbles. From figure~\ref{curvature}, we notice that the cusp radius 
of  curvature scales  like ${\tau}^2$. 
  Notice that the smallest radius we could measure is around 200 $\mu m$. Although it is 
  expected that the cusp curvature increases above this value, we do not have enough 
  video resolution to go beyond this value at which numerical noise becomes 
  important. The inset of figure~\ref{curvature} shows how fast the radius 
  of curvature drops in time and is indeed a signature of a finite time
  singularity process.

If we suppose that, before the cusp is developed, the only forces present in the system are the pressure 
drop at the hole and the pressure difference across the interface, then the 
scaling can be self consistently derived as the following:
The divergence in capillary stress  is balanced by a divergence in the 
viscous stress at the exit hole and which scales as $\beta \mu Q/(dh)^3$ \cite{hbrenner}, 
where $Q$ is the flux of the fluid out  of the drain hole, $\beta$ is a number of the order of unity
and $\mu$ is the viscosity.  Here we are assuming that 
the flow associated with the dimple formation is a sink flow which is verified experimentally
by monitoring the flow far and close to the hole by monitoring the flow using PIV (Particle 
Image Velocimetry) and we used passive tracers such as small bubbles to monitor the displacement of a volume element of the liquid.
The air bubble we followed is  around 200 $\mu m$ in diameter with a rise velocity smaller than 10 $\mu$m/sec measured by tracking
the bubble path. Since the flow at 
the hole has a velocity of 1 -2 mm/sec, we suppose that the ascending velocity 
of the tracer bubble is 
negligible over the time scale of the observation and the bubble follows essentially the 
flow.

\begin{figure}[htb!]
\centerline{\epsfysize=\columnwidth \epsfbox{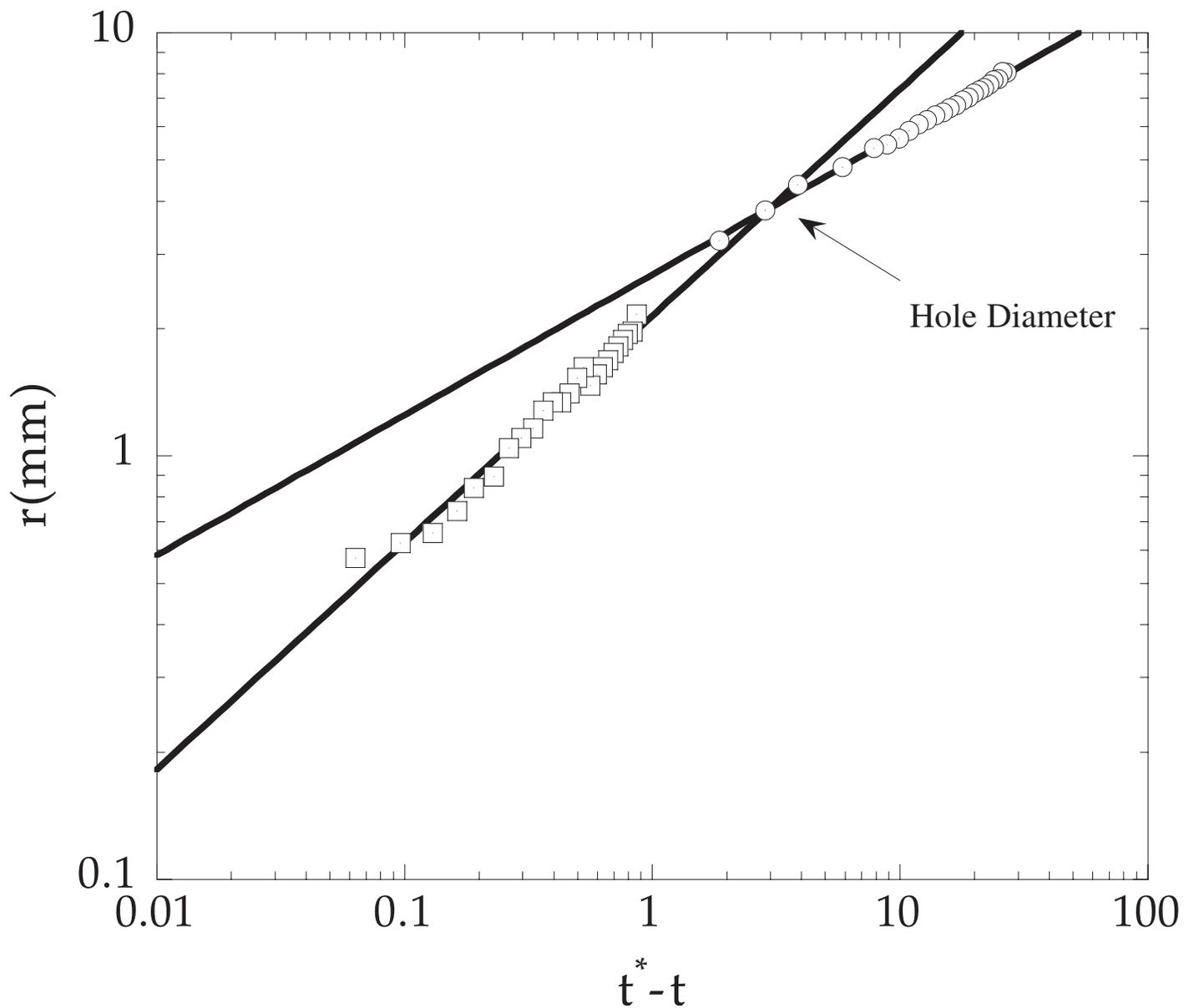}}
\caption{Position of a passive tracer versus the time elapsed between the 
onset of the draining and the time $t^*$ when the tracer disappears in the hole 
of the container and the coordinates origin ($r=0$)
corresponds to the center of the hole at the level of the bottom of the 
plate. The lines are  1/3 power law (open circles) and 
 1/2 power law (opened squares). The crossover between these two 
tendencies lies at the hole diameter. These data points were obtained for 
the same run.}
\label{position}
\end{figure}

In figure~\ref{position}, we depict the position $r$ of the tracer versus 
the time before it reaches the hole. Here $r$ is the position of a point in the fluid where the 
origin $r=0$ is at the center of the hole at the level of the bottom 
surface and $t^*$ is the time at which the tracer passes through the 
hole. We plot $r$ versus $t^* -t$. The fit to the plot in  figure ~\ref{position}, shows 
that the particle's position scales like $(t^*-t)^{.33}$ for long
times and far  away from the hole, and it scales like $(t^*-t)^{.5}$ for short times or closer to the hole.
 The scaling of $r$ versus $t^*-t$ closer to the hole can be explained by
writing the equation for a stokes flow, knowing that the pressure drop
at the hole is given by the poiseuille formula which gives the expression of the pressure
gradient along the hole and the flow rate, the geometry and the fluid properties \cite{poiseuille}. 
  
From this equation  \cite{poiseuille} we know that the only quantity on which $v$ can depend, other than $z$ and $t$ is $\nu$. 
From these three quantities only one dimensionless combination can be found which is $ \eta = z/(\nu t)^2$.
From this scaling we note that the vertical distance $z$  goes like $t^{1/2}$. This explain the behavior
$r \propto (t^* -t)^{1/2}$.

We  interpret the second result with a simple (heuristic) argument as the 
following:  When the tracer
is far from the singularity, the velocity field is the due to a dipole, since
the singularity has an image across the horizontal plate, due to the 
no-slip condition. We know from 
low reynolds number hydrodynamics \cite{hbrenner} that the dipolar contribution to 
the velocity field, due to a sink flow, 
is such that $v \sim 1/r^2$, and since the tracer is passive as we pointed 
it out earlier in the text, the velocity is simply dr/dt, solving in t, we find that
$r \sim (t^{*}-t)^{1/3}$, which is the behavior found experimentally. 
This proves experimentally that the flow in the tank is indeed a sink flow. This
property will allow us to use the results of sink flow mainly the
pressure drop at the sink which is given by  $\beta \mu Q/(dh)^3$.
 We will suppose our process as quasi-static and we can 
balance capillary forces and viscous forces due to the sink flow at every step of the drainage.
From the balance $\gamma \kappa \approx \beta \mu Q/(dh)^3$ we see that close to the 
formation of the cusp, $\kappa \approx 1/(dh)^3$. Measurements of $\kappa$ 
and $dh$ show that they have the following power law scaling with the time 
remaining before the formation of the cusp: $\kappa \propto (t_f -t)^{-2}$ 
and $dh \propto (t_f - t)^{2/3}$.  Although the scaling is explained self-consistently,  
it is unclear at present what determines 
the values of these exponents.
These exponents and the curvature drop in the inset of 
Fig.~\ref{curvature}, are a signature of finite time singularity.

In summary, we have examined the  motion of an air-liquid interface due 
to drainage at very low reynolds numbers where viscosity is dominant. We found that the prior to cusp formation, the interface 
height and the dimple curvature follow scaling laws dictated by the interplay 
between surface tension and viscous pressure at the drainage hole.

SC thanks Martine Benamar, Howard Brenner, 
Michael Brenner, Denis Gueyffier  for helpfull discussions and G. H. McKinley for his support. This 
work was initiated during the year 1999-2000 and supported by a NASA grant NAG3-2155.


\begin{thebibliography}{10}
\bibitem{poiseuille}
The pressure gradient along the $z$ axis of a cylinder
if given by: $\frac{\partial P}{\partial z} = \frac{8\mu Q}{\pi D^4}$
where the $z$ axis is the vertical axis along gravity, $Q$ is the flow rate, 
$\mu$ is the viscosity and $D$ is the radius of the cylinder. The Navier-Stokes
equation becomes $\frac{\partial v}{\partial t} = \frac{8\nu Q}{\pi D^4} + 
\nu \frac{\partial ^2 v}{\partial z^2}$.
\end{thebibliography}
\bibliographystyle{unsrt}

\end{document}